\documentclass[epj]{svjour}
\usepackage{graphics}
\usepackage{amssymb}
\usepackage{color}
\begin{document}
\title{Axiomatic nonextensive statistics at NICA energies}
\author{Abdel Nasser Tawfik\inst{1,2,3 }
\thanks{http://atawfik.net/}%
}                     
%
%
\institute{Egyptian Center for Theoretical Physics (ECTP), Modern University for Technology and Information (MTI), 11571 Cairo, Egypt \and World Laboratory for Cosmology And Particle Physics (WLCAPP), 11577 Cairo, Egypt \and 
Academy for Scientific Research and Technology (ASRT), Network for Nuclear Sciences (NNS), 11511 Cairo, Egypt}
\date{Received: 15 October 2015 / Revised version: 21 January 2016}
%
\abstract{
We discuss the possibility of implementing axiomatic nonextensive statistics, where it is conjectured that the phase-space volume determines the (non)extensive entropy, on the particle production at NICA energies. Both Boltzmann-Gibbs and Tsallis statistics are very special cases of this generic (non)extensivity. We conclude that the lattice thermodynamics is {\it ab initio} extensive and additive and thus the nonextensive approaches including Tsallis statistics categorically are not matching with them, while the particle production, for instance the particle ratios at various center-of-mass energies, is likely a nonextensive process but certainly not of Tsallis type. The resulting freezeout parameters, the temperature and the chemical potentials, are approximately compatible with the ones deduced from Boltzmann-Gibbs statistics.
\PACS{
      {05.30.-d}{ Quantum statistical mechanics}   \and
      {25.75.Dw}{ Particle production in relativistic collisions} \and
      {02.50.Cw}{ Probability theory}
     } 
} 
\maketitle
\section{Introduction}

Besides beam extraction from the nuclotron, the existing accelerator of heavy ions at the Joint Institute for Nuclear Research (JINR), the future Nuclotron based Ion Collider fAcility (NICA) foresees the construction of a collider and three detectors; multi-purpose detector (MPD), baryonic matter at nuclotron (BM@N) and spin physics detector (SPD). NICA is designed to cover a wide range of the quantum choromodynamic (QCD) phase-diagram at the highest baryonic density. The nucleon-nucleon center-of-mass energies ($\sqrt{s_{NN}}$) at NICA cover the range from $4$ up to $11~$GeV at luminosity $\sim 10^{27}~$ cm$^{-2}\cdot$ s$^{-1}$ for Au$^{79+}$, at which a maximal net-baryon (freezeout) density shall be formed. This makes NICA optimum for the exploration of compressed nuclear matter and reveal essential properties of QCD at high density. Furthermore, the various colliding ions available in this energy range and the high luminosity enable NICA verify novel aspects of the high-energy collisions. 

In the present work, we discuss the theoretical ideas about the nonextensive statistics and its role in characterizing the power scaling and the statistical nature of the particle production at finite temperature and high density. Section \ref{sec:nonextensv} is devoted to the various proposals for the nonextensive statistics at high density. An axiomatic nonextensive partition-function shall be presented in section \ref{sec:pfunction}. Implications on lattice QCD thermodynamics and reproduction of various particle ratios shall be discussed in the successive two sections, respectively. The final conclusions are elaborated in section \ref{sec:cons}.

\section{Nonextensive statistics at high density}
\label{sec:nonextensv}

In a system consisting of two isolated subsystems ($A$ and $B$), where $\Omega_A$ and $\Omega_B$ are the respective numbers of states, extensive entropy is guaranteed if $S(\Omega_{A+B})=S(\Omega_A)+S(\Omega_B)$. The additivity is defined as $S(\Omega_A\, \Omega_B)=S(\Omega_A)+S(\Omega_B)$ \cite{Thurner1,Thurner2}. Both extensivity and additivity coincide if $\Omega_{A+B} =\Omega_A\, \Omega_B$ \cite{Thurner1}, which is obviously fulfilled for Boltzmann-Gibbs (BG) statistics; $S[p] = \sum_{i}^{\Omega} g(p_i)$, where  $g(p_i)=-p_i\, \ln p_i$. 

In a recent work \cite{TE2015}, we have shown that the blind implementation of Tsallis algebra (replacing exponential and logarithmic functions with their Tsallis counter parts) \cite{Tsallis1988,PratoTsallis1999,Tsallis2005} in the partition-function of the hadron resonance gas (HRG) model, which is - per definition - constructed from summing up independent contributions from different hadron resonances, fails to assure full incorporation of nonextensivity (due to correlations or interactions, among others) in such statistical thermal models. The resulting HRG is no longer able to reproduce the lattice QCD calculations, and furthermore results in very low temperatures even from the statistical fitting with the transverse momentum spectra ($p_T$), which are exceptionally (but seemingly unjustly) celebrated as the best implication for the Tsallis statistics in high-energy physics \cite{worku1,Cleymans2014}. 

The resulting chemical freezeout phase-diagram is very different from the one deduced from {\it extensive} statistical models \cite{Tawfik:2014eba}. The latter fits well with various recent lattice calculations \cite{Depman2015} and different experimental measurements. For example, the freezeout temperatures determined from the thermal ensemble in which Tsallis statistics is implemented, are much smaller than the temperatures characterizing the {\it extensive} freezeout \cite{Depman2015}.

Concretely, dor Tsallis-statistics, the resulting nonextensive temperature, which considerably has been interpreted as follows \cite{BiroBook}.
\begin{itemize}
\item First, due to fundamental differences between BG- and Tsallis-statistics and hence an extrapolation of the {\it ordinary} $T_{ch}$ has been proposed \cite{deppman2012}
\begin{eqnarray}
T_q &=& T_{ch} + (q-1)\, k,
\end{eqnarray}
where constant $k$ is modelled to be dependent on the {\it so-called} energy transfer between the source and the surrounding.  At $T_{ch}=192\pm 15~$MeV, $c=-(950\pm 10)~$MeV \cite{sena}.
\item Second, based on a physical model of finite thermostat \cite{Biro,BiroEntropy,Biro2015}
\begin{eqnarray}
T_{ch} &=& T_q\, \exp\left(-\frac{S_{q}}{C}\right),
\end{eqnarray}
where $C$ is finite heat reservoir capacity, which together with the fluctuations in $T$  modify the canonical exponential weight factor \cite{Biro2015}. It is assumed that $C$ determines whether $q$, which shall be elaborated later on,  is smaller or larger than unity.
\end{itemize}
It is obvious that such {\it phenomenological} interpretations are closely related to the blind, {\it improper} implementation of Tsallis-algebra. Furthermore, even the excellent reproduction of the transverse momentum distributions seems to be misinterpreted \cite{bialas2015}, and accordingly, the resulting temperature.

Recently, Bialas explained the incompleteness of the claim that the transverse momentum distribution of different produced particles at high energy is to be reproduced (actually fitted) from statistical models employing Tsallis algebra \cite{bialas2015}. First, the contradiction between the proposed applicability of such statistical models at high energy and the perturbative QCD, which likely dominates such an energy region, remains unsolved. Second, the decay of statistical clusters can be scaled as power laws very close to the ones from the Tsallis statistics. Such power laws seem to exist over a wide range of transverse momenta. Therefore, the decay of statistical clusters might explain the surprising agreement between the measured transverse momenta and the calculations from statistical models with Tsallis algebra. In other words, the power-law scaling might be stemming from the decay of statistical cluster rather than the Tsallis-nonextensivity.

We show that the nonextensivity of statistical thermodynamics plays an essential role in describing various aspects of high-energy physics, especially at high density. NICA precise measurements considerably improve the certainties of the experimental quantities, which shall be utilized in exacting the nonextensivity of the particle production, that likely accompanying critical phenomena, such as the phase transitions and/or the critical endpoint and/or the out-of-equilibrium processes. These quantities include the particle yields and the particle ratios, the transverse momenta, the higher-order moments, etc. 

The assumption that the high-energy collision goes through several successive processes, is now very widely accepted. Most of them, for instance the hadronization processes, are strongly correlated and radically changing their properties, symmetries and degrees of freedom \cite{Thurner2}. Such an ensemble violates one of the four Shannon-Khinchin axioms \cite{ShannonKhinchin}. Therefore, the entropic nonextensivity should be fully assured, where the Tsallis entropy is a very special type.  In light of this, we mention two ingredients.
\begin{itemize}
\item First, long time ago, a pioneering nonextensive generalization to the Hagedorn's approach was proposed \cite{beck1}. An interesting result on the agreement between the temperature fluctuations and the high-energy scattering measurements was reported \cite{beck2}. Recently, superstatistics rather than Tsallis was introduced as the relevant nonextensive approach for high-energy physics \cite{beck3}. 
\item Second, it was shown that Tsallis statistics likely violates CPT symmetry \cite{tamas}. From Kubo-Martin-Schwinger (KMS) relation \cite{kms}, a kind of generic quantum statistical distributions has been proposed, in which the exponential function fulfilling that $\exp_{\kappa}(x) \cdot \exp_{\kappa}(-x)=1$, should be re-expressed as 
\begin{eqnarray}
\exp_{\kappa}(x) &=& \left\{\left[1+ (b\, k\, x)^2\right]^{1/2} + b\, k\, x\right\}^{1/k}, \\
b &=& \frac{\left[\exp_q(x)\right]^k - \left[\exp_q(-x)\right]^k}{2\, k\, x},
\end{eqnarray}
known as $\kappa$-statistics for constant $b$, only. $q$ differing from unity characterizes Tsallis statistics, $k=1/(1-q)$ and $q$-exponential function $\exp_q(-x)$ \cite{Tsallis1988,PratoTsallis1999,Tsallis2005}.  For an arbitrary $b$, a more general form for the quantum statistical distribution should be obtained.   
We have evaluated $\kappa$-statistics and noticed (not shown here) that the numerical estimation for the actual implementation of the logarithmic function is similar to the one proposed by Tsallis algebra.  But the resulting $T$ considerably differs from the Boltzmannian one. It is also true that this fit parameter is not the same when $p_T$ or $m_T$ or $m_T-m$, or the Lorentz-boosted $m_T-m$ distributions are taken into consideration. We conclude that $\kappa$-algebra actually does not considerably improve the confrontation with the high-energy and the lattice thermodynamics, for instance, relative to Tsallis.  
\end{itemize}

\subsection{Entropies leading to non-exponential canonical distributions}

Here, we shortly list out various types of entropies leading to  non-exponential canonical distributions \cite{Thurner2,Biro2015}. 
\begin{itemize}
\item Renyi entropy is a generalisation of Boltzmann-Gibbs-Shannon entropy \cite{renyi1961}:
\begin{eqnarray}
S_R &=& \frac{1}{1-q} \ln \sum_i^{\Omega} p_i^q,
\end{eqnarray}
where $p_i$ is the probability of $i$-th state and the parameter $q$ possesses various physical meanings.
\item Tsallis nonextensive entropy \cite{Tsallis1988}:
\begin{eqnarray}
S_T &=& \frac{1}{1-q} \ln \sum_i^{\Omega} (p_i^q - p_i).
\end{eqnarray}
\item Generalized entropies assuring extensivity and thus the possibility of considering interacting statistical systems, section \ref{sec:axiomatic}  \cite{Thurner2}:
\begin{eqnarray}
S_q[p] &=& \sum_{i=1}^{\Omega} g(p_i),
\end{eqnarray}  
where the degree of nonextensivity in the function $g$, which can have any functional form, for instance incomplete gamma function, is restrictively determined by the four Khinchin axioms \cite{ShannonKhinchin}.
\end{itemize}

\subsection{Correspondence of phase space and  nonextensivity in high-energy collisions} 
\label{sec:phasespc}

As discussed in Ref. \cite{TE2015}, a great discrepancy between the extensive ($q=1$) and the nonextensive ($q\neq1$, i.e. Tsallis) treatment for high-energy collisions and the lattice QCD calculations is observed. It was proposed that this could be interpreted due to the differences between the power laws imposed by Tsallis statistics and the Boltzmann factor. Despite the remarkably low temperature deduced from the statistical fitting of Tsallis algebra and the transverse momentum spectra, 
\begin{itemize}
\item an {\it ad hoc} factorization was imposed to the Tsallis-statistics. This factorization is only valid in the system of interest that can be divided into micro states, and down to single states with single particles and even so it was implemented in a mathematically improper way. 
\item Even the mathematically correct factorization \cite{factrz} is nothing but an approximative approach to the exact Tsallis-statistics. In other words, Tsallis-statistics describes a special case in a general way and what is so far implemented in high-energy and lattice thermodynamics is an approximation to that statistics.
\item  Last but not least, a mean value for $q$ was considered instead of the standard average. The Tsallis statistics should be implemented to micro or even single states and accordingly, $q_i$ should be first assigned to each $i$-th micro or single state. Out of this, one has to estimate the standard average in order to replace the mean value used so far.
\end{itemize}

The question is how the statistics which describes the phase-space volume, the number of states and the elementary cells should be modified when describing correlations among non-Boltzmannian subsystems? In case of locally correlated, equal and distinguishable $\Omega$ subsystems, the extensive Boltzmann-Gibbs entropy can be utilized \cite{GellMannTsallis2005},
\begin{itemize}
\item if such subsystems become globally correlated, a vast class of entropies can be implemented, where $q\neq 1$ is one special case, 
\item if these subsystems become noncorrelated (independent), Boltzmann-Gibbs entropy becomes additive as well, and 
\item in case that all marginal probabilities of discrete binary-subsystems have been reached, an asymptotic scale-freedom is obtained. Such duality likely fixes a unique escort probability \cite{ThurnerGellMann2012}.
\end{itemize}
It was concluded \cite{ThurnerGellMann2012} that the dependence of the complexity of generalized entropy on the phase space might not entirely be given by the simple binary correlations.

\subsection{Distribution function and axiomatic entropy}
\label{sec:axiomatic}

For a large statistical system, the generalized entropies can be classified according to their asymptotic properties \cite{Thurner1}. Two classes have been proposed \cite{Thurner1}. To each of them, a scaling function is assigned, which is characterized by the exponents $c$ or $d$ for first or second property, respectively, 
\begin{eqnarray}
S_{c,d}[p] &=& \sum_{i=1}^{\Omega} A\, \Gamma(d+1, 1 - c \log p_i) - B\, p_i, \label{eq:NewExtns1}
\end{eqnarray}
where $\Gamma (a, b)=\int_{b}^{\infty}\, dt\, t^{a-1} \exp (-t)$ is incomplete gamma-function, and $A$ and $B$ are arbitrary parameters. In the limit that $\Omega \rightarrow \infty$, then each of the asymptotic subsystems can be described by Eq. (\ref{eq:NewExtns1}). The universality class $(c, d)$ does not only characterize the entropy in a complete way (whether extensive or nonextensive), but also specifies the correspondent distribution function \cite{Thurner1},
\begin{eqnarray}
\varepsilon_{c,d,r}(x) = \exp\left\{\frac{-d}{1-c}\left[\mathtt{W}_k\left(B\left(1-x/r\right)^{1/d}\right) - \mathtt{W}_k(B)\right]\right\}, \hspace*{3mm} \label{eq:ps1}
\end{eqnarray}
where $\mathtt{W}_k$ is the $k$-th Lambert-{$\mathtt{W}$} function, which has real solutions at $k=0$ for all classes with $d\geq 0$ and also at $k=−1$ for negative $d$. 
\begin{eqnarray}
B &\equiv & \frac{(1-c)r}{1-(1-c)r}\, \exp \left[\frac{(1-c)r}{1-(1-c)r}\right],
\end{eqnarray} 
with $r=(1-c+c d)^{-1}$. At $k=0$, Lambert-{$\mathtt{W}$} function can asymptotically be expanded as 
\begin{eqnarray}
{\mathtt{W}}(x) &=& \sum_{n=1}^{\infty} \frac{(-1)^{n-1}\, n^{n-2}}{(n-1)!}\, x^n.
\end{eqnarray}

As discussed in Ref. \cite{Thurner1}, the properties of this generic extensivity condition, Eq. (\ref{eq:NewExtns1}), lead to
\begin{eqnarray}
\left(1-c\right)^{-1} &=& \lim_{N \rightarrow \infty} N\, \frac{\Omega^{\prime}}{\Omega}, \label{5} \\
d &=& \lim_{N \rightarrow \infty} \log \Omega\, \left(\frac{1}{N} \frac{\Omega}{\Omega^{\prime}}+c-1\right). \label{6}
\end{eqnarray}
The number of micro states ($\Omega$) can be related to the distribution function, itself, 
\begin{eqnarray}
\Omega(N) &=& 
  \exp \left\{\frac{d}{1-c} \mathtt{W}_k \left[\frac{(1-c)\, e^{\frac{1-c}{c\, d}}}{c\, d} \left( \frac{\varphi\, c\, N}{r} \right)^{1/d}\right] \right\} \nonumber \\
  &\times & \frac{1}{\varepsilon_{c,d}(-\varphi\, c\, N)}, \label{eq:states1}
\end{eqnarray}
where $\varphi$ is given as 
\begin{eqnarray}
\varphi = \frac{d}{d N}\, S_g &=& \Omega^{\prime}\, \left[g(1/\Omega)-\frac{1}{\Omega}\, g^{\prime} (1/\Omega)\right]. \label{scd}
\end{eqnarray}

\subsection{Axiomatic nonextensive statistics and hadronization processes}

In the quark coalescence model, the produced hadrons can be estimated by valence quarks and/or antiquarks multiplied by the coalescence coefficient $C_{hadron}$ and a nonlinear normalization coefficient $b_{q}$. The latter takes into consideration the conservation of the various quantum numbers \cite{c3}. 
\begin{eqnarray}
N_M &=&D^M\, C_M(i,j)\, b_{q_{i}}\, N_{q_{i}}\, b_{\overline{q_{j}}}\, N_{\overline{q_{j}}},  \label{eq:mesonNumber}\\
N_B &=& D^B\, C_B(i,j,k)\, b_q(i)\, N_q(i)\; b_q(j)\, N_q(j)\, b_q(k)\, N_q(k). \hspace*{5mm} \label{eq:baryonNumber}
\end{eqnarray}
In obtaining these expressions, it was assumed that the hadrons are produced in a thermal equilibrium \cite{c44} and the spin degeneracy ($S_{hadron}$) and $D^{hadron}=2\, S_{hadron}+1$ are additional normalization factors. Straightforwardly, one goes from nonlinear to linear hadronization processes. The hadron number becomes directly proportional to the quark constituents multiplied by  the reaction volume \cite{c5}, for instance, 
\begin{eqnarray}
N_p  &\propto & N_q^{3}, \\
N_{\Lambda|\Sigma}  &\propto & N_q^{2}\, N_s, \\
N_\Xi  & \propto & N_q\,  N_s^{2}, \\
N_{\Omega}  & \propto & N_s^{3}.
\end{eqnarray}
The proportionality factor and the exponents of the quark flavors give an estimation for the extensivity parameter $d$ \cite{TE2015}.
 
The coalescence model is proposed because: 
\begin{enumerate}
\item it implies a special grouping of the valence quarks forming baryons and mesons,
\item it very well describes the hadronization process in high-energy collisions, and 
\item thus constant and proportional connectivity and constant connectancy can be characterized and accordingly the values of $c$ and $d$ can be guessed. 
\end{enumerate}

Assuming constant connectivity, which minimizes the possibilities of forming hadrons from a concrete number of quarks, then from Eqs. (\ref{eq:mesonNumber}) and (\ref{eq:baryonNumber}),  the available quark numbers ($N_q$) determine the number of the resulting hadrons,
\begin{eqnarray}
N_M \propto  \frac{1}{2}\, N_q,  & \qquad &
N_B  \propto  \frac{1}{3}\, N_q. \label{eqLNMB}
\end{eqnarray}
Thus, the possible hadron states can be estimated as
\begin{eqnarray}
\Omega &=& \left(\begin{array}{c} N_q \\ N_{M|B}\end{array}\right),
\end{eqnarray}
which can be re-expressed as a function of either $N_q$ or $N_{M|B}$. Accordingly, we get $c=d=1$, i.e. Boltzmann-Gibbs extensive entropy.

There are three possibilities:
\begin{itemize}
\item Constant connectivity. Then Eq. (\ref{eqLNMB}) leads to $d\, N_{N|B}\propto \alpha\, N_q$, with $\alpha={\cal O}(2)$ for bosons and $\alpha={\cal O}(3)$ for baryons. Hence, the maximum number of possible hadron states  reads
\begin{eqnarray}
\Omega & \simeq & b^{N_q},
\end{eqnarray} 
and the statistical system can be described by extensive entropy, i.e.  $c=d=1$. 
\item Constant connectancy. This leads to $k\propto N_q$ and accordingly, 
\begin{eqnarray}
\Omega & \simeq & N_q^{N_{M|B}}. 
\end{eqnarray}
The statistical system can then be described by nonextensive $q$-entropy, i.e. Tsallis statistics. 
\item Proportional connectivity which represents a departure from the main assumption of simplest confining procedure. In this case $k\propto N_q^{\alpha}$ leads to a super-equipotential increase in the entropy, i.e. $c=1$ and $d=1/\alpha_s$, where $\alpha_s$ is the strong running coupling. 
\end{itemize}

\section{Axiomatic nonextensive partition-function}
\label{sec:pfunction}

At vanishing baryon chemical potential and from Eq. (\ref{eq:ps1}), the partition function of and axiomatic nonextensive ensemble is given as
\begin{eqnarray}
\ln\, Z_{cl}(T) &=& V\, \sum_i^{N_{M|B}}\, g_i \int_0^{\infty} \frac{d^3 p}{(2 \pi)^3}\; \varepsilon_{c,d,r}(x_i), \label{eq:PFcdr1}
\end{eqnarray}
where $V$ being the fireball volume and  $x_i$ stands for the $i$-th resonance dispersion relation, $x_i=(p^2+m_i^2)^{1/2}$, normalized to temperature $T$. 

For Fermi-Dirac and Bose-Einstein quantum statistics, 
\begin{eqnarray}
\ln\, Z(T) &=& \pm V\, \sum_i^{N_{M|B}}\, g_i \int_0^{\infty} \frac{d^3 p}{(2 \pi)^3}\; \ln\left[1\pm\varepsilon_{c,d,r}(x_i)\right],  \hspace*{4mm} \label{eq:PFcdr2}
\end{eqnarray}
where $\pm$ stands fermions and bosons, respectively. All thermodynamic quantities can be derived from Eq. (\ref{eq:PFcdr1}) or Eq. (\ref{eq:PFcdr2}).

\subsection{Confronting extensive and nonextersive statistics to lattice QCD thermodynamics}

\begin{figure}[htb]
\centering{
\resizebox{0.5\textwidth}{!}{
\includegraphics{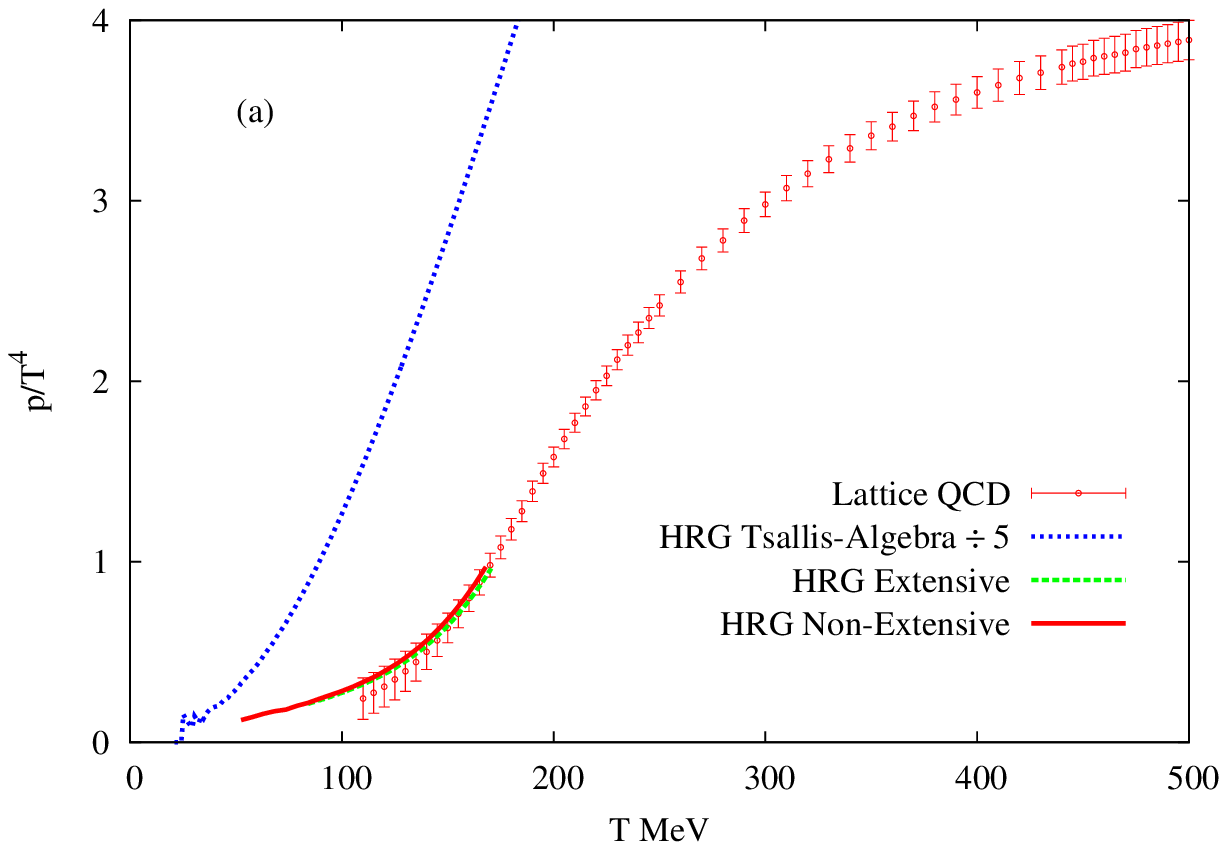}
\includegraphics{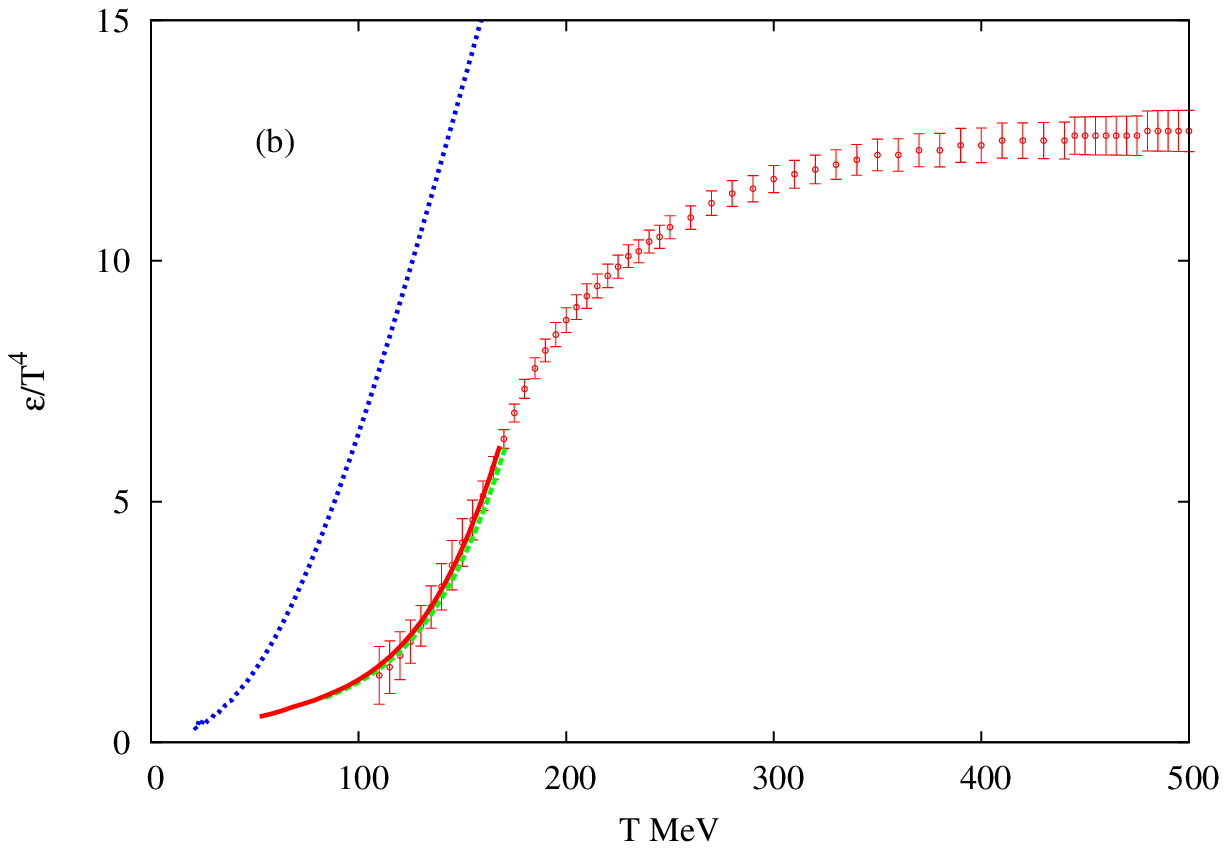}
}
\caption{Left-hand panel (a) presents a comparison between lattice QCD  pressure ($p/T^4$) \cite{latticeQCD} (symbols) and the calculations from statistics thermal models (HRG) with extensive (dashed), nonextersive axiomatic (solid) and Tsallis algebra (dotted curve). The right-hand panel (b) depicts the same as in left-hand panel but for the energy density ($\epsilon/T^4$). \label{fig:fg2}}
}
\end{figure}

Instead of the {\it ad hoc} assumption that $c=0$, while assigning Tsallis nonextensive parameter $q$ to $d$, we apply Eq. (\ref{eq:PFcdr2}), which takes into consideration various possibilities of extensivity and nonextensivity. The results on  pressure $p/T^4$ (left-hand panel) and energy density $\epsilon/T^4$ (right-hand panel) are depicted in Fig. \ref{fig:fg2}. Both thermodynamic quantities are estimated at a vanishing chemical potential (related to very high center-of-mass energies), at which the lattice QCD simulations become very reliable. Despite the sign-problem making Monte-Carlo techniques no longer applicable, $\mu_b/T\leq 1$ sets limits to the applicability of the lattice calculations. At larger $\mu_b$ (comparable to NICA), we are left with QCD-like approaches. This explains the reason, why we limit to discussion here to reliable lattice calculations, only.

It is obvious that the calculations from the statistical thermal models, for instance HRG, in which the nonextensivity is defined from the axiomatic entropy, reproduce very well the lattice QCD calculations. In this approach, the two asymptotic properties, $c$ and $d$, to each of them a scaling function  is associated, are implemented. This defines a generalized entropy for such strongly correlated systems. In order to reproduce the lattice simulations, values very close to unity should be assigned to both scaling exponents $c$ and $d$ (solid curves). These lead to the conclusion that the classical Boltzmann-Gibbs statistics reproduces very well the lattice results. To confirm this conclusion, we also present calculations from extensive HRG (dashed curves), i.e. fully Boltzmann-Gibbs approach ($c=d=1$), and found that the lattice results are also very well reproduced (dashed curves). The results from Tsallis statistics are depicted as dotted curves. It is obvious that even when normalizing them to the factor $5$, the lattice calculations can not be reproduced.  So far, we conclude that the lattice QCD thermodynamics is likely extensive.

\subsection{Particle ratios from extensive and axiomatic nonextersive statistics}

\begin{figure}[htb]
\centering{
\resizebox{0.5\textwidth}{!}{
\includegraphics{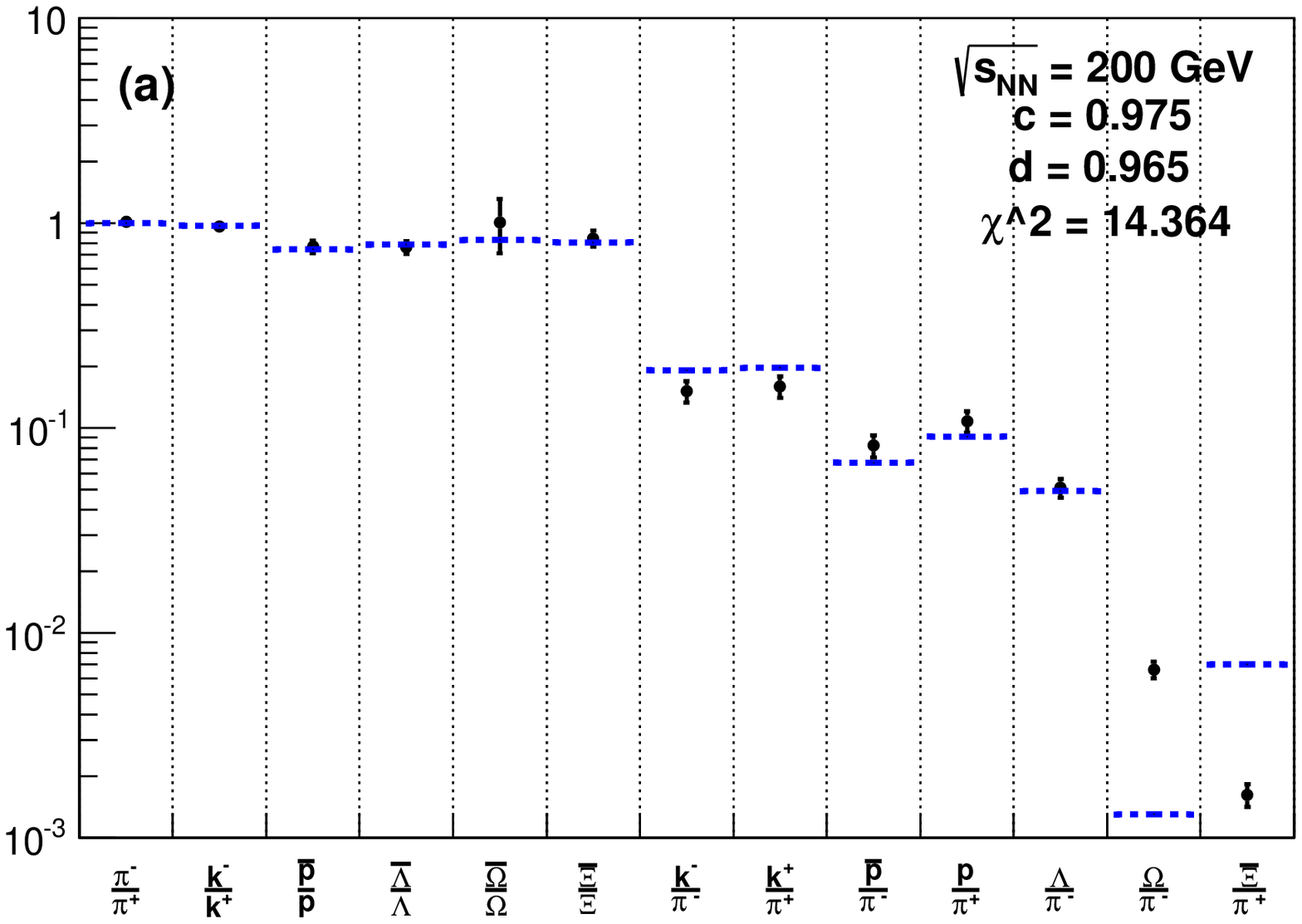}
\includegraphics{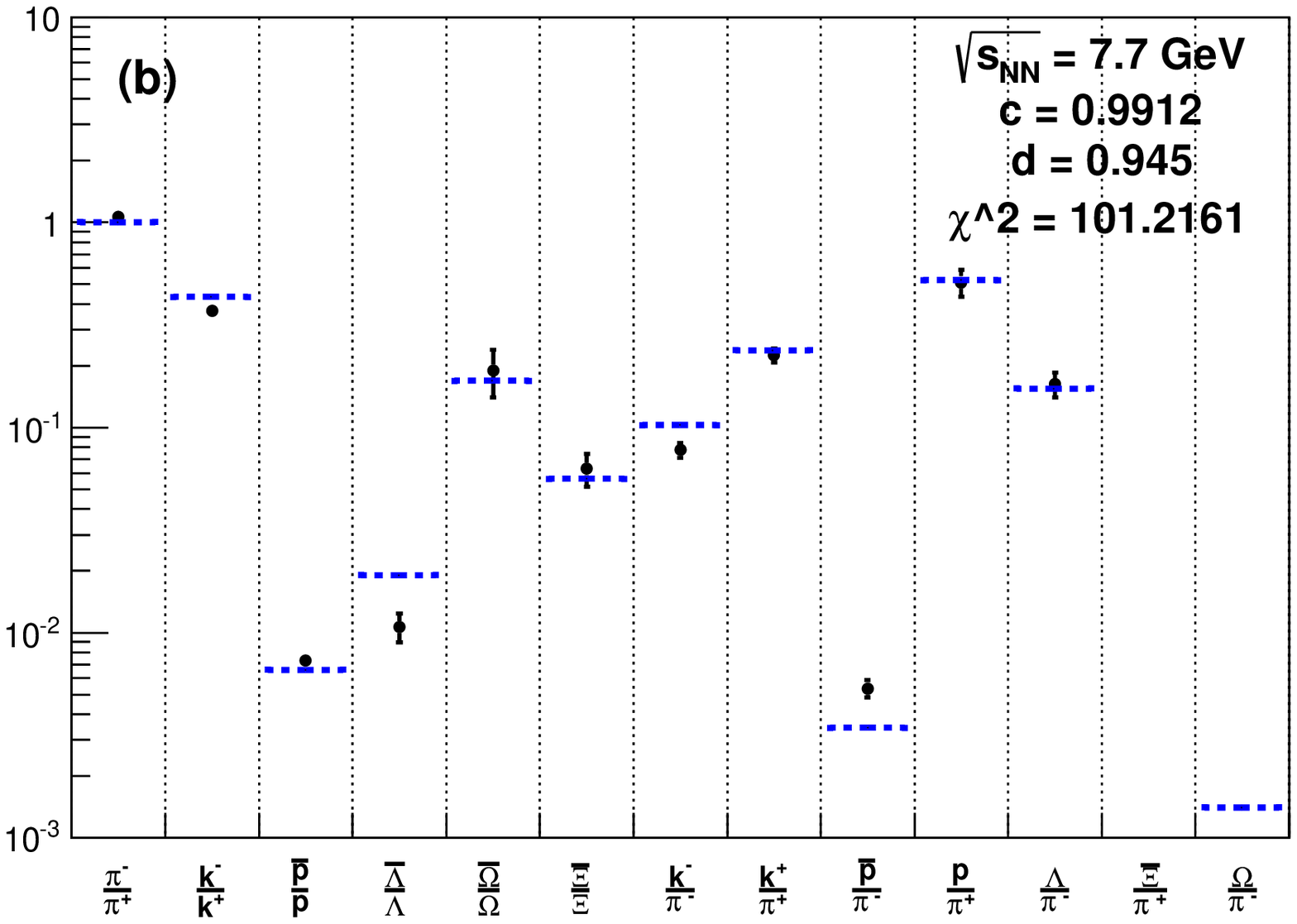}
} 
\caption{Left-hand pane (a): different particle ratios deduced from axiomatic nonextensive statistics (dashed lines), Eq. (\ref{eq:PFcdr2}), are compared with the experimental measurements at $200~$GeV (symbols). Right-hand panel (b) shows the same but at $7.7~$GeV. \label{fig:prnonext}}
}
\end{figure}

Examples on different particle ratios measured at $200~$GeV [left-hand panel (a)] and at $7.7~$GeV [right-hand panel (b)] are depicted in Fig. \ref{fig:prnonext}. The experimental measurements are compared with the statistical fitting from the proposed axiomatic nonextensive statistics (dashed lines). At $200~$GeV, $c=0.975$ and $d=0.965$ while at $7.7~$GeV, which is very compatible with NICA energies, $c=0.9912$ and $d=0.945$. The quality of both fittings defers at these two respective energies; $\chi^2/dof=1.105$ and $\chi^2/dof=7.785$. It is worthwhile to notice that the discrepancy with the experimental results is mainly in reproducing strange particles. Their fugacity factors likely play a crucial role. Also, the quark occupation factor and the excluded-volume effects might be responsible for such discrepancy. This will be a subject of a future study. The resulting freezeout parameters read
\begin{itemize}
\item at $200~$GeV, $T_{ch}=148.05~$MeV and $\mu_b=23.94~$MeV and
\item at $7.7~$GeV, $T_{ch}=145.32~$MeV and $\mu_b=384.3~$MeV,
\end{itemize}
which are obviously very compatible with the ones deduced from Boltzmann-Gibbs statistics \cite{Tawfik:2013bza}. We conclude that the resulting $c$ and $d$ refer to neither Boltzmann-Gibbs- nor Tsalis-type statistics. But in order to confirm whether $c$ increases and/or $d$ reduces with the energy, a further analysis should be conducted. It seems that at $7.7~$GeV, which belongs to NICA energy-range, $c\simeq 1$, while $d<1$. Again, both values neither match with Boltzmann-Gibbs nor with Tsallis statistics.

\section{Conclusions}
\label{sec:cons}

We conclude that the lattice calculations {\it ab initio} assume extensivity and likely additivity, as well. Therefore, all nonextensive approaches including Tsallis statistics categorically are not suitable to reproduce them. The proposed axiomatic nonextensivity at $c=d\simeq 1$, i.e. extensive Boltzmann-Gibbs statistics, simulates very well the lattice QCD thermodynamics.

The particle production is likely a nonextensive process, i.e. both $c$ and $d$ differ from unity. On the other hand, this process can not be described by the very special nonextensive Tsallis statistics. This opens a wide horizon to analyse particle yields, ratios, fluctuations, higher-order moments, etc. by means of this generic nonextensive approach. NICA precise measurements shall play a central role.

%

\end{document}